\documentclass[twocolumn,trackchanges]{aastex7}

\accepted{to PASP on \today}

\usepackage{amsmath}
\usepackage{mathtools}
\usepackage{enumitem}

\newcommand{\JPL}{Jet Propulsion Laboratory, California Institute of Technology, 4800 Oak Grove Drive, Pasadena, CA 91109, USA}

\newcommand{\UCO}{UC Observatories, University of California, Santa Cruz, CA 95064, USA}

\newcommand{\WMKO}{W.\ M.\ Keck Observatory, 65-1120 Mamalahoa Hwy, Kamuela, HI 96743, USA}
\newcommand{\SSL}{Space Sciences Laboratory, University of California, Berkeley, CA 94720, USA}

\newcommand{\UCB}{Department of Astronomy, 501 Campbell Hall, University of California, Berkeley, CA 94720, USA}
\newcommand{\UCLA}{Department of Physics \& Astronomy, University of California Los Angeles, Los Angeles, CA 90095, USA}
\newcommand{\nexsci}{NASA Exoplanet Science Institute/Caltech-IPAC, California Institute of Technology, Pasadena, CA
91125, USA}
\newcommand{\COO}{Caltech Optical Observatories, California Institute of Technology, Pasadena, CA 91125, USA}

\newcommand{\Schmidt}{Astrophysics \& Space Institute, Schmidt Sciences, New York, NY 10011, USA}

\begin{document}

\title{A Hot Jupiter with a Retrograde Orbit around a Sun-like Star \\ and a Toy Model of Hot Jupiters in Wide Binary Star Systems}

\correspondingauthor{Steven Giacalone}

\author[0000-0002-8965-3969]{Steven Giacalone}
\altaffiliation{NSF Astronomy and Astrophysics Postdoctoral Fellow}
\affiliation{Department of Astronomy, California Institute of Technology, Pasadena, CA 91125, USA}
\email[show]{giacalone@astro.caltech.edu}

\author[0000-0001-8638-0320]{Andrew W. Howard}
\affiliation{Department of Astronomy, California Institute of Technology, Pasadena, CA 91125, USA}
\email{ahoward@caltech.edu}

\author[0000-0003-3856-3143]{Ryan A. Rubenzahl}
\affiliation{Center for Computational Astrophysics, Flatiron Institute, 162 Fifth Avenue, New York, NY 10010, USA}
\email{rrubenzahl@flatironinstitute.org}

\author[0000-0002-8958-0683]{Fei Dai}
\affiliation{Institute for Astronomy, University of Hawai`i, 2680 Woodlawn Drive, Honolulu, HI, 96822, USA}
\email{fdai@hawaii.edu}

\author[0000-0002-9305-5101]{Luke B. Handley}
\affiliation{Department of Astronomy, California Institute of Technology, Pasadena, CA 91125, USA}
\email{lhandley@caltech.edu}

\author[0000-0002-0531-1073]{Howard Isaacson}
\affiliation{\UCB}
\email{hisaacson@berkeley.edu}

\author[0000-0003-1312-9391]{Samuel Halverson}
\affil{\JPL}
\email{samuel.halverson@jpl.nasa.gov}

\author[0009-0008-9808-0411]{Max Brodheim}
\affiliation{\WMKO}
\email{mbrodheim@keck.hawaii.edu}

\author{Matt Brown}
\affiliation{\WMKO}
\email{mbrown@keck.hawaii.edu}

\author[0000-0001-6416-1274, sname=Carmichael, gname=Theron]{Theron W. Carmichael}
\altaffiliation{NSF Ascend Fellow}
\affiliation{Institute for Astronomy, University of Hawai`i, 2680 Woodlawn Drive, Honolulu, HI, 96822, USA}
\email{tcarmich@hawaii.edu}

\author[0009-0000-3624-1330]{William Deich}
\affil{\UCO}
\email{will@ucolick.org}

\author[0000-0003-3504-5316]{Benjamin J.\ Fulton}
\affiliation{\nexsci}
\email{bjfulton@ipac.caltech.edu}

\author[0009-0004-4454-6053]{Steven R. Gibson}
\affiliation{\COO}
\email{sgibson@caltech.edu}

\author[0000-0002-7648-9119]{Grant M.\ Hill}
\affiliation{\WMKO}
\email{ghill@keck.hawaii.edu}

\author[0000-0002-6153-3076]{Bradford Holden}
\affil{\UCO}
\email{holden@ucolick.org}

\author[0000-0002-5812-3236]{Aaron Householder}
\affiliation{Department of Earth, Atmospheric and Planetary Sciences, Massachusetts Institute of Technology, Cambridge, MA 02139, USA}
\affil{Kavli Institute for Astrophysics and Space Research, Massachusetts Institute of Technology, Cambridge, MA 02139, USA}
\email{aaron593@mit.edu}

\author[0000-0003-2451-5482]{Russ R. Laher}
\affiliation{NASA Exoplanet Science Institute/Caltech-IPAC, 1200 E California Blvd, Pasadena, CA 91125, USA}
\email{laher@ipac.caltech.edu}

\author[0009-0004-0592-1850]{Kyle Lanclos}
\affiliation{\WMKO}
\email{klanclos@keck.hawaii.edu}

\author[0009-0008-4293-0341]{Joel Payne}
\affiliation{\WMKO}
\email{jpayne@keck.hawaii.edu}

\author[0000-0003-0967-2893]{Erik A. Petigura}
\affiliation{\UCLA}
\email{petigura@astro.ucla.edu}

\author[0000-0001-8127-5775]{Arpita Roy}
\affiliation{\Schmidt}
\email{arpita308@gmail.com}

\author[0000-0002-4046-987X]{Christian Schwab}
\affiliation{School of Mathematical and Physical Sciences, Macquarie University, Balaclava Road, North Ryde, NSW 2109, Australia}
\email{mail.chris.schwab@gmail.com}

\author[0009-0007-8555-8060]{Martin M.\ Sirk}
\affiliation{\SSL}
\email{sirk@ssl.berkeley.edu}

\author[0000-0002-6092-8295]{Josh Walawender}
\affiliation{\WMKO}
\email{jwalawender@keck.hawaii.edu}


\begin{abstract}

We report an observation of a transit of the hot Jupiter (HJ) KELT-23A b with the Keck Planet Finder spectrograph and a measurement of the sky-projected obliquity ($\lambda$) of its Sun-like ($T_{\rm eff} \approx 5900$~K) host star. We measured a projected stellar obliquity of $\lambda \approx 180^\circ$, indicating that the orbit of the HJ is retrograde relative to the direction of the stellar spin. Due to the slow sky-projected rotational velocity of the host star ($v \sin{i_\star} \approx 0.5$~km~s$^{-1}$), the true orbit of the HJ could be closer to polar. HJs around stars with effective temperatures below the Kraft break -- such as KELT-23A -- are generally found to have prograde orbits that are well-aligned with the equatorial planes of their host stars (i.e., $\lambda \sim 0^\circ$), most likely due to spin-orbit realignment driven by stellar tidal dissipation. This system is therefore a unique outlier that strains migration and tidal theories. The fact that the HJ has a highly misaligned orbit may suggest that the planet arrived at its close-in orbit relatively recently, possibly via interactions with the wide-separation (570 AU) M-dwarf companion in the system, or that it has stalled near an antialigned or polar orientation while realigning. Using {\it Gaia} DR3, we determined the orbit of the stellar companion to be moderately face-on ($\gamma = 60 \pm 4^\circ$). We show that the distribution of observed systems in the $\gamma - \lambda$ plane can be broadly reproduced using a toy model in which the orbits of the planetary and stellar companions begin aligned with the equatorial plane of the primary star and, upon migrating inwards, the planet preferentially obtains either an aligned or polar orbit.

\end{abstract}



\section{Introduction} \label{sec:intro}

Hot Jupiters (HJs; Jupiter-like planets orbiting close to their host stars) are generally believed to form far from their stars and move inwards via one of two mechanisms: migration through the protoplanetary disk or high-eccentricity tidal migration, the process by which the orbit of the planet is excited into a highly eccentric orbit that subsequently shrinks and circularizes due to tidal interaction with the star near periastron (see \citealt{dawson2018origins} for a review). While the existence of these close-in giants has been known for decades, the dominant mechanism responsible for their observed orbits has long remained a mystery.

Stellar obliquity -- the angle between the stellar spin axis and the orbital axis of its transiting planetary companion -- contains unique clues about the orbital evolutions of planets. In general, disk migration should primarily result in planet orbits that are roughly coplanar with the equators of their host stars (hereafter referred to as ``aligned'' orbits) whereas high-eccentricity tidal migration should often result in planets orbits that are significantly inclined relative to the equators of their host stars (hereafter referred to as ``misaligned'' orbits). An informative summary of the stellar obliquity distributions resulting from different eccentricity-excitation mechanisms can be found in \citet{albrecht2022obluqities}. 

The sky-projected component of the stellar obliquity ($\lambda$), accessible via the Rossiter-McLaughlin effect \citep{mclaughlin1924, rossiter1924}, has become a particularly valuable tool for understanding the origins of HJs \citep[e.g.,][]{rice2022origins, dong2023obliquity, siegel2023obliquity}. \citet{winn2010obliquities} identified the first trends in the $\lambda$ distribution for stars with HJs, finding that these stars tend to have low $\lambda$ (i.e., aligned orbits) when $T_{\rm eff} < 6250$~K and often have high $\lambda$ (i.e., misaligned orbits) when $T_{\rm eff} > 6250$~K. This dichotomy has primarily been interpreted as the consequence of more efficient tidal dissipation around stars cooler than the Kraft break, where stars have radiative interiors and convective exteriors \citep{albrecht2012obliquities}. Because HJs around hotter stars tend to have misaligned orbits, many have argued that high-eccentricity migration offers the most compelling explanation for their close-in orbits.

In this connection, HJs in binary star systems are notable case studies of inward migration. Wide-separation stellar companions are believed to be capable of driving a number of mechanisms that result in misaligned orbits \citep[e.g.,][]{anderson2016kz, yang2025hatp7}, including some in which the protoplanetary disk itself is forced into a highly inclined orientation \citep[e.g.,][]{batygin2012primordial}. Given the higher probability of hotter stars belonging to multi-star systems \citep{moe2017mind}, these mechanisms may be partially responsible for the observed $\lambda$ distribution if they are highly efficient. Over the years, the number of stars with both transiting planets and wide binary stars with measured $\lambda$ has steadily grown, allowing for the first investigations of the relationship between the orbits of the planetary and stellar companions \citep{behmard2022companions, rice2024geometries}.

In this paper, we expand this sample with an obliquity measurement of the star KELT-23A, a Sun-like star ($T_{\rm eff} = 5899 \pm 49$~K, $R_\star = 0.996 \pm 0.015 \, R_\odot$, $M_\star = 0.944^{+0.060}_{-0.054} \, M_\odot$) with a wide-separation binary M dwarf companion and a transiting HJ on a 2.26-day orbit \citep{johns2019kelt23}. The paper is organized as follows. In Section~\ref{sec:observations}, we describe the observations used in our analysis of the KELT-23 system. In Section~\ref{sec:analysis}, we detail our analysis of the data. In Section~\ref{sec:discussion}, we place the KELT-23 system into context by comparing it to other systems. In addition, we present a toy model that can broadly reproduce the observed orientations of systems with both HJs and wide-separation stellar companions. We provide concluding remarks in Section~\ref{sec:conclusion}.

\section{Observations} \label{sec:observations}

\subsection{Keck Planet Finder Spectroscopy}

We observed a transit of KELT-23A~b on 24 April 2024 UT using the high-resolution ($R \sim 97,000$) Keck Planet Finder (KPF) spectrograph on the 10-meter Keck-I telescope to derive precise radial velocities of the host star \citep{gibson2024kpf}. Our observations were scheduled via the KPF Community Cadence project \citep{lubin2025astroQ}. We acquired 32 480-second exposures between 9:40 and 14:23 UT with the standard 47-second detector readout, achieving peak signal-to-noise ratios (SNRs) per pixel of 180 in the green arm ($445-600$~nm) and 220 in the red arm ($600-870$~nm) after stacking the three science traces. These observations covered the full transit of the HJ, which took place between 10:40 and 13:00 UT, and approximately 1.5 hours of out-of-transit baseline. At the beginning and end of the sequence, we acquired a single wavelength calibration exposure using a Fabry–Pérot etalon in order to track and correct for intranight instrumental drift \citep{schwab2015etalon}. The data were reduced and radial velocities were derived using the publicly available KPF pipeline.\footnote{\url{https://github.com/Keck-DataReductionPipelines/KPF-Pipeline}} The KPF radial velocities (RVs) are shown in Table~\ref{tab:data}.

\begin{deluxetable}{ccc}
\tablewidth{0pt}
\tablecaption{KPF Radial Velocity Data \label{tab:data}}
\tablehead{
\colhead{Time (BJD)} & \colhead{RV (m~s$^{-1}$)} & \colhead{$\sigma_{\rm RV}$ (m~s$^{-1}$)}
}
\startdata
$ 2460424.90737 $ & $ -16953.06 $ & $ 1.15 $ \\
$ 2460424.91354 $ & $ -16955.25 $ & $ 1.16 $ \\
$ 2460424.91962 $ & $ -16957.36 $ & $ 1.17 $ \\
$ 2460424.92580 $ & $ -16959.35 $ & $ 1.17 $ \\
$ 2460424.93182 $ & $ -16961.95 $ & $ 1.16 $ \\
$ 2460424.93804 $ & $ -16965.27 $ & $ 1.17 $ \\
$ 2460424.94416 $ & $ -16968.31 $ & $ 1.14 $ \\
$ 2460424.95008 $ & $ -16971.74 $ & $ 1.16 $ \\
$ 2460424.95635 $ & $ -16977.42 $ & $ 1.19 $ \\
$ 2460424.96264 $ & $ -16981.47 $ & $ 1.18 $ \\
$ 2460424.96864 $ & $ -16982.71 $ & $ 1.22 $ \\
$ 2460424.97468 $ & $ -16988.47 $ & $ 1.22 $ \\
$ 2460424.98084 $ & $ -16984.77 $ & $ 1.23 $ \\
$ 2460424.98694 $ & $ -16988.93 $ & $ 1.21 $ \\
$ 2460424.99312 $ & $ -16987.96 $ & $ 1.25 $ \\
$ 2460424.99916 $ & $ -16990.45 $ & $ 1.17 $ \\
$ 2460425.00528 $ & $ -16992.65 $ & $ 1.23 $ \\
$ 2460425.01144 $ & $ -16994.04 $ & $ 1.23 $ \\
$ 2460425.01742 $ & $ -16995.77 $ & $ 1.26 $ \\
$ 2460425.02376 $ & $ -16997.55 $ & $ 1.30 $ \\
$ 2460425.02975 $ & $ -17004.45 $ & $ 1.27 $ \\
$ 2460425.03595 $ & $ -17007.00 $ & $ 1.30 $ \\
$ 2460425.04212 $ & $ -17010.01 $ & $ 1.27 $ \\
$ 2460425.04819 $ & $ -17013.93 $ & $ 1.25 $ \\
$ 2460425.05416 $ & $ -17016.76 $ & $ 1.23 $ \\
$ 2460425.06034 $ & $ -17018.99 $ & $ 1.25 $ \\
$ 2460425.06654 $ & $ -17021.45 $ & $ 1.25 $ \\
$ 2460425.07282 $ & $ -17024.00 $ & $ 1.24 $ \\
$ 2460425.07882 $ & $ -17025.82 $ & $ 1.18 $ \\
$ 2460425.08472 $ & $ -17029.31 $ & $ 1.23 $ \\
$ 2460425.09108 $ & $ -17032.31 $ & $ 1.34 $ \\
$ 2460425.09708 $ & $ -17034.72 $ & $ 1.28 $
\enddata
\end{deluxetable}



\subsection{\textit{TESS} Photometry}

We also utilized time-series photometry of the KELT-23 system from the Transiting Exoplanet Survey Satellite (\textit{TESS}; \citealt{ricker2010transiting}) in the joint fit. \textit{TESS} observed the system for a total of 19 sectors, each of which spans roughly 27 days in length. Data from each sector was collected as a 2-minute cadence and was analyzed by the \textit{TESS} Science Processing Operations Center (SPOC; \citealt{jenkins2016spoc}). Data from seven of the sectors were also collected at a 20-second cadence. The full \textit{TESS} data set begins in July of 2019 and ends in October of 2024, spanning more than five years.

\begin{deluxetable*}{lllll}
\tablewidth{0pt}
\tablecaption{Definitions, priors, and best-fit values of the free parameters utilized in the joint \texttt{DYNESTY} fit \label{tab:results}}
\tablehead{
\colhead{Parameter} & \colhead{Description} & \colhead{Units} & \colhead{Prior} & \colhead{Best-Fit Value}
}
\startdata
\multicolumn{5}{c}{Shared Model Parameters}\\
\hline
$T_{0}$ & Transit epoch  & TBJD & $\mathcal{U}(2389.815,2389.817)$ & $2389.816371 \pm 0.000016$ \\
$P_{\rm orb}$ & Orbital period  & days & $\mathcal{U}(2.25,2.26)$ & $2.255287666 \pm 0.000000051$ \\
$R_{\rm p} / R_\star$ & Planet-star radius ratio & --- & $\mathcal{U}(0.1,0.2)$ & $0.13300 \pm 0.00026$ \\
$a/R_\star$ & Semi-major axis over stellar radius& --- & $\mathcal{U}(5,10)$ & $7.615 \pm 0.021$ \\
$\cos{i}$ & Cosine of the orbital inclination & --- & $\mathcal{U}(0.00,0.25)$ & $0.06884 \pm 0.00088$ \\
\hline
\multicolumn{5}{c}{Transit Model Parameters}\\
\hline
$q_{1}$ & {\it TESS} quadratic limb darkening coeff. 1 & --- & $\mathcal{U}(0,1)$ & $0.350 \pm 0.029$ \\
$q_{2}$ & {\it TESS} quadratic limb darkening coeff. 2 & --- & $\mathcal{U}(0,1)$ & $0.138 \pm 0.051$ \\
$\gamma_{\rm T}$ & {\it TESS} flux baseline & --- & $\mathcal{U}(-1,1)$ & $0.000201 \pm 0.000014$\\
$\ln{\sigma_\mathrm{s, {\rm T}}}$ & Natural logarithm of {\it TESS} scatter & --- & $\mathcal{U}(-20,0)$ & $-8.350^{+0.085}_{-0.100}$ \\
\hline
\multicolumn{5}{c}{R-M Model Parameters}\\
\hline
$\lambda$ & Sky-projected stellar obliquity & degrees & $\mathcal{U}(0,360)$ & $180.4^{+4.9}_{-4.7}$ \\
$v \sin{i_\star}$ & Sky-projected stellar rotational velocity & km s$^{-1}$ & $\mathcal{U}(0,5)$ & $0.468^{+0.044}_{-0.043}$ \\
$\cos{i_\star}$ & Cosine of the stellar inclination & --- & $\mathcal{U}(0,1)$ & $0.51 \pm 0.34$ \\
$u_{1}$ & KPF quadratic limb darkening coeff. 1 & --- & $\mathcal{U}(0,1)$ & $0.63^{+0.23}_{-0.28}$ \\
$u_{2}$ & KPF quadratic limb darkening coeff. 2 & --- & $\mathcal{U}(0,1)$ & $0.57^{+0.29}_{-0.33}$ \\
$\gamma_{\rm K}$ & KPF radial velocity baseline & m s$^{-1}$ & $\mathcal{U}(-16980,-16920)$ & $-16952.29 \pm 0.45$ \\
$\dot{\gamma}_{\rm K}$ & KPF radial velocity slope & m s$^{-1}$ day$^{-1}$ & $\mathcal{U}(-500,-400)$ & $-434.65^{+3.65}_{-3.69}$ \\
$\ln{\sigma_\mathrm{K}}$ & Natural logarithm of KPF uncertainty & $\ln{\rm m \, s^{-1}}$ & $\mathcal{U}(-20,10)$ & $-0.07^{+0.15}_{-0.13}$ \\
\hline
\multicolumn{5}{c}{Derived Quantities}\\
\hline
$i$ & Orbital inclination & degrees & --- & $86.052^{+0.051}_{-0.050}$ \\
$b$ & Transit impact parameter & --- & --- & $0.5243 \pm 0.0053$ \\
\enddata
\tablecomments{Best-fit values and their uncertainties are the 16th, 50th, and 84th percentiles of the posterior distributions. For the priors, $\mathcal{U}$(l,u) represents a uniform distribution with lower and upper bounds of l and u, respectively. TBJD is the {\it TESS} BJD, defined as TBJD = BJD - 2457000.}
\end{deluxetable*}

\begin{figure*}[t!]
    \centering
    \hspace{-0.5cm}\includegraphics[width=0.48\linewidth]{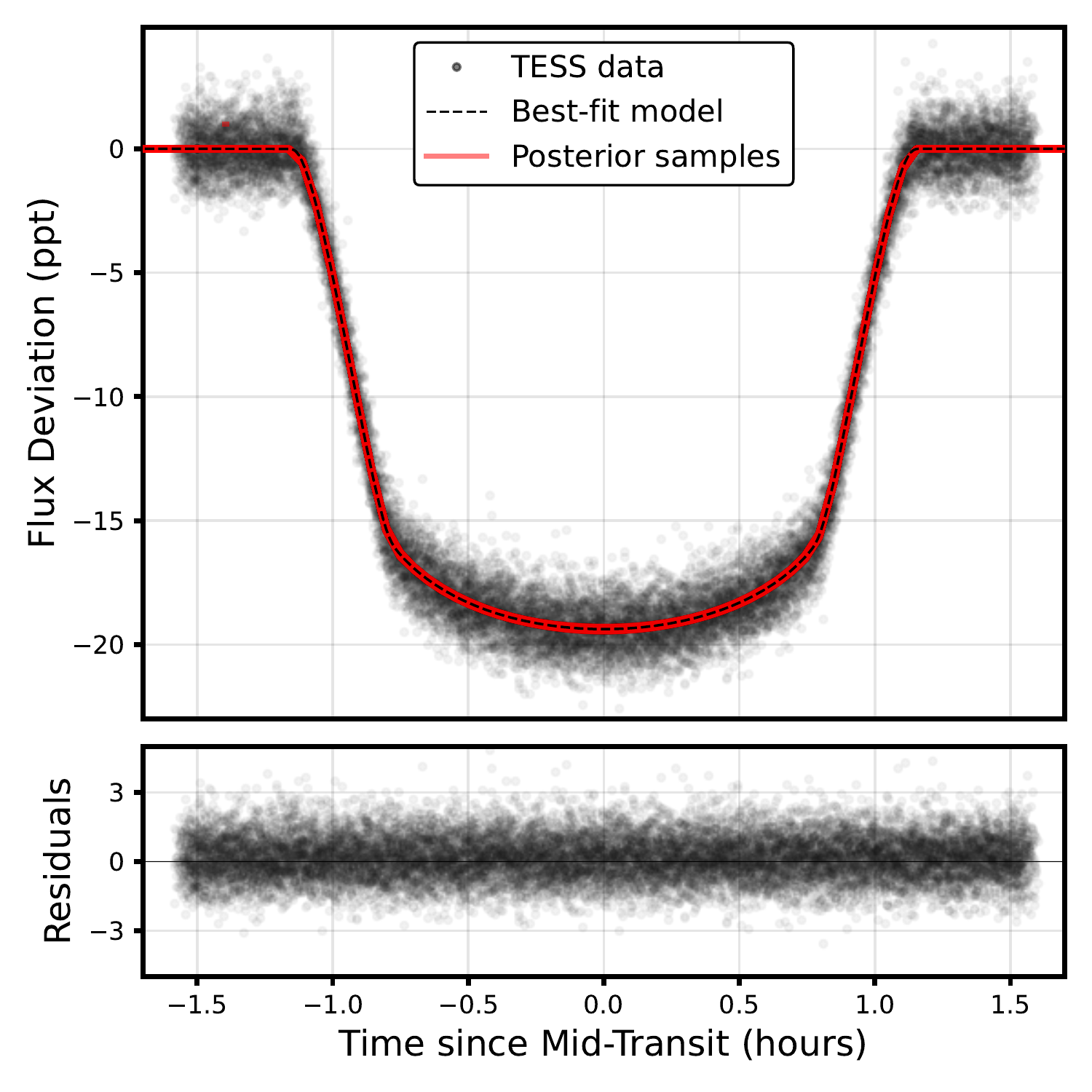}
    \includegraphics[width=0.48\linewidth]{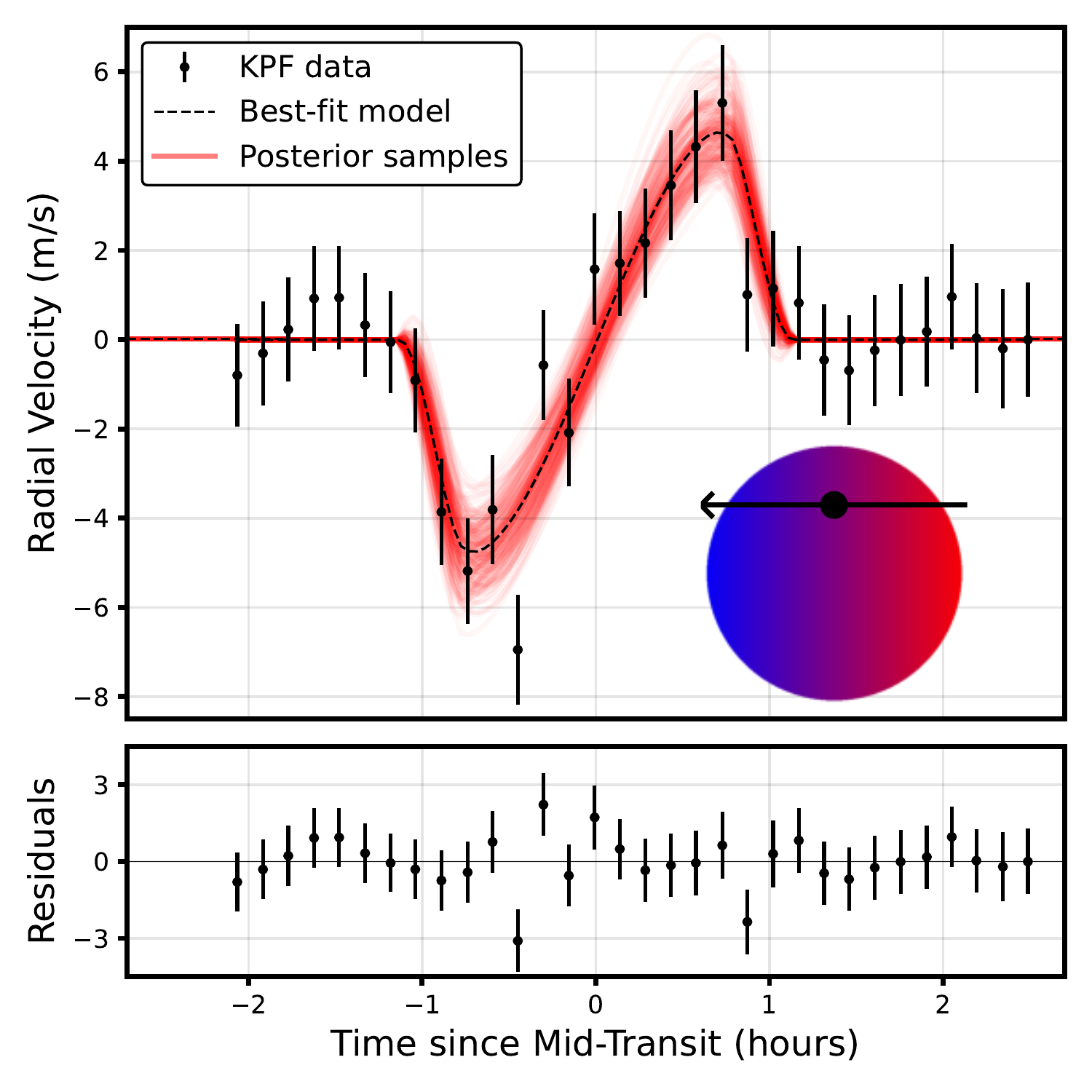}
    \caption{Results of the joint \texttt{DYNESTY} fit to the {\it TESS} photometry (left) and KPF radial velocities (right). A diagram of the best-fit transit is shown in the right-hand panel, where the color of the stellar surface indicates whether it is blueshifted and redshifted. The black points are the data, the dashed black lines are the best-fit (median) models, and the solid red lines are random samples from the posterior distributions (200 in total). The {\it TESS} data is phase folded to the best-fit period and transit epoch. The offset and slope have been removed from the radial velocity data for visual clarity.}
    \label{fig:fit}
\end{figure*}

\section{Analysis} \label{sec:analysis}

\subsection{Stellar Obliquity}

To measure the projected stellar obliquity ($\lambda$) of the system, we jointly fit the {\it TESS} photometry and KPF radial velocity data using \texttt{DYNESTY}, a dynamic nested sampler implemented in Python \citep{speagle2020dynesty}. The {\it TESS} data were modeled using the \texttt{BATMAN} Python package \citep{kriedberg2015batman} and the radial velocity data were modeled using the equations in \citet{hirano2011rm}, which solve for the anomalous radial velocity arising from the R-M effect. Following the adopted orbital parameters in \citet{johns2019kelt23}, we assumed a circular orbit. For the joint fit, we utilized a joint likelihood function of the form
\begin{equation}
    \ln{L_{\rm tot}} = \ln{L_{\rm T}} + \ln{L_{\rm K}} 
\end{equation}
where the likelihood of a model given the data is Gaussian in nature. For the {\it TESS} data, the likelihood takes the form
\begin{equation}
        \ln{L_{\rm T}} = - \frac{1}{2} \sum_i \left[ \frac{(y_{{\rm T},i} - f(t_{{\rm T},i}; \theta_{\rm T}))^2}{\sigma_{{\rm T},i}^2} - \log{\sigma_{{\rm T},i}^2}\right].
\end{equation}
Here, $t_{{\rm T},i}$, $y_{{\rm T},i}$, and $\sigma_{{\rm T},i}$ are the time, flux, and flux error of the $i^{\rm th}$ data point, respectively. To account for additional astrophysical and instrumental noise in the {\it TESS} data, we define $\sigma_{{\rm T},i}^2 = \sigma_{m,{\rm T},i}^2 + \sigma_{s,{\rm T}}^2$, where $\sigma_{m,{\rm T},i}$ is the measurement uncertainty of the data point and $\sigma_{s, {\rm T}}$ is a systematic error term for which we fit. For the KPF data, the likelihood takes the form
\begin{equation}
        \ln{L_{\rm K}} = - \frac{1}{2} \sum_i \left[ \frac{(y_{{\rm K},i} - f(t_{{\rm K},i}; \theta_{\rm K}))^2}{\sigma_{{\rm K}}^2} - \log{\sigma_{{\rm K}}^2}\right].
\end{equation}
Again, $t_{{\rm K},i}$ and $y_{{\rm K},i}$ are the time and radial velocity of the $i^{\rm th}$ data point. For the radial velocity model, we do not include the individual radial velocity measurement uncertainties and instead treat $\sigma_{\rm K}$ as a free parameter. This is because the current KPF pipeline tends to overestimate measurement uncertainties, making it impossible to constrain radial velocity jitter (the equivalent of $\sigma_{s,{\rm K}}$) in our case. 

The model parameters ($\theta$) for the transit and radial velocity models are defined in Table~\ref{tab:results} and include five shared free parameters, four free parameters that belong to only the transit model, and eight free parameters that belong only to the radial velocity model. Each of the free parameters is assigned a uniform prior. The models for the two datasets are denoted by $f(t_i, \theta)$. For the transit model, this function is
\begin{equation}
    \begin{multlined}
        f_{\rm T}(t_{{\rm T},i}; \theta_{\rm T}) = \\
        {\rm BM}(t_{{\rm T},i}; T_0; P_{\rm orb}; R_{\rm p}/R_\star; a/R_\star; i; q_{\rm 1}; q_{\rm 2}) + \gamma_{\rm T}  
    \end{multlined}
\end{equation}
where BM is the output of \texttt{BATMAN} and $\gamma_{\rm T}$ is a flux offset term. For the radial velocity model, this function is
\begin{equation}
    \begin{multlined}
        f_{\rm K}(t_{{\rm K},i}; \theta_{\rm K}) = \\
        {\rm RM}(t_{{\rm K},i}; T_0; P_{\rm orb}; R_{\rm p}/R_\star; a/R_\star; i; \lambda; v\sin{i_\star}; i_\star; u_{\rm 1}; u_{\rm 2}) \\ 
        + \gamma_{\rm K} + \dot{\gamma}_{\rm K}(t_i - t_0)  
    \end{multlined}
\end{equation}
where RM is the output of the R-M model defined in \citet{hirano2011rm}, $\gamma_{\rm K}$ is a radial velocity offset term, $\dot{\gamma}_{\rm K}$ is a radial velocity slope term, and $t_0$ is the time of the first KPF observation. The radial velocity offset and slope terms are intended to estimate the reflex motion of the star due to the planet and any other companions in the system as well as stellar activity, which are approximately linear over the timescale of our observation for inactive stars. In the R-M model, we assumed a microturbulent velocity of 0.7~km~s$^{-1}$, a macroturbulent velocity of 4.18~km~s$^{-1}$ (solved for using Equation~1 of \citet{valenti2005spocs} with $T_{\rm eff} = 5899$~K), a natural line width of 1~km~s$^{-1}$, and an instrumental Gaussian dispersion of 1.56~km~s$^{-1}$ (the KPF line-spread function width). Lastly, we note that we ignored the effects of differential rotation and convective blueshift. The effects of differential rotation are negligible at small $v\sin{i_\star}$. Convective blueshift is known to cause radial velocity signals of magnitudes of $\sim 2$~m~s$^{-1}$, but often generally does not impact the measured value of $\lambda$ significantly \citep[e.g.,][]{rubenzahl2021wasp107, handley2025toi1694}.\footnote{When including convective blueshift in the model, we recovered an obliquity of $\lambda = 181.8^{+5.8}_{-5.6}$$^\circ$, which agrees well with that recovered by the convective-blueshift-free model, and a largely unconstrained convective blueshift velocity of $v_{\rm cb} = -233^{+166}_{-148}$~m~s$^{-1}$}.

The \texttt{DYNESTY} sampler was initialized with the default settings, which included 500 live points, multiple bounding ellipsoids for prior bound approximation, and a random walk sampling of the likelihood space. The sampler was terminated when the condition $\log{(z + z_r) - \log{z}} < 0.01$ was satisfied, where $z$ is the current evidence from all saved samples and $z_r$ is the estimated contribution to the evidence from the remaining likelihood volume. The results of this fit are shown in Table~\ref{tab:results} and Figure~\ref{fig:fit}. 


We measured a sky-projected stellar obliquity of $\lambda = 180.4^{+4.9}_{-4.7}$$^\circ$, indicating an orbit that is retrograde relative to the stellar spin. We also precisely measured the sky-projected stellar rotational velocity from the R-M fit to be $v \sin{i_\star} = 0.468^{+0.044}_{-0.043}$~km~s$^{-1}$, which is relatively low for a Solar-type star (note that a uniform prior of $0-5$~km~s$^{-1}$ was assumed for $v\sin{i_\star}$ in the fit).\footnote{We note that the corner plot between $\lambda$ and $v \sin{i_\star}$ revealed no noticeable covariance between the two variables.} This may suggest a nearly face-on stellar spin axis. Based on the analysis in \citet{johns2019kelt23}, the system has an age of $6.4^{+3.5}_{-3.2}$~Gyr, for which Sun-like stars typically exhibit rotation periods of 20--40~days \citep{barnes2007gyro, mamajek2008age, angus2015gyro}. For KELT-23A, these rotation periods correspond to stellar inclinations between roughly $10^\circ$ and $20^\circ$ assuming a stellar radius of $R_\star = 0.996 \pm 0.015 \, R_\odot$ \citep{johns2019kelt23}. This range of stellar inclination angles would translate to true 3-dimensional stellar obliquities between roughly $95^\circ$ and $110^\circ$, or a near-polar orbit \citep{masuda2020stellarinc}. We note that a similar argument for a polar orbit was made by \citet{winn2009hatp7} for HAT-P-7, another multi-star system with a HJ for which $\lambda \sim 180^\circ$.

We emphasize that this is a purely speculative exercise. It is possible that the spin of KELT-23A has been slowed due to tidal interactions with its close-in HJ (see Section~\ref{sec:discussion} for a more detailed discussion). Ultimately, a measurement of the stellar rotation period is required to calculate the actual stellar inclination and 3D stellar obliquity. While it is possible to measure the rotation period of a star based on starspot-driven modulation of its light curve, we found no strong evidence of strong modulation in the \textit{TESS} data.



\subsection{The Orbit of KELT-23B}

Following \citet{behmard2022companions}, we used {\it Gaia}~DR3 to precisely measure the relative positions and velocity vectors of the two stars KELT-23A and KELT-23B \citep[also see][]{tokovinin2016eccentricity}. The angle between the relative position vector and relative velocity vector, typically denoted $\gamma$, contains information about the orbit of the pair of stars around their center of mass. For instance, $\gamma = 0^\circ$ suggests that we are viewing the system edge-on whereas $\gamma = 90^\circ$ suggests that we are viewing the system face-on (assuming a near-circular orbit). When combined with the measured obliquity of the primary star, this angle can reveal information about the origin and evolution of the system \citep[e.g.,][]{behmard2022companions, rice2024geometries}.


For the KELT-23 system, we measured $\gamma = 60 \pm 4^\circ$, suggesting that the pair of stars orbit one another closer to a face-on orientation than an edge-on orientation. We further discuss the implications of this angle in Section~\ref{sec:discussion}.


\begin{figure*}[t!]
    \centering
    \hspace{-0.6cm}\includegraphics[width=\linewidth]{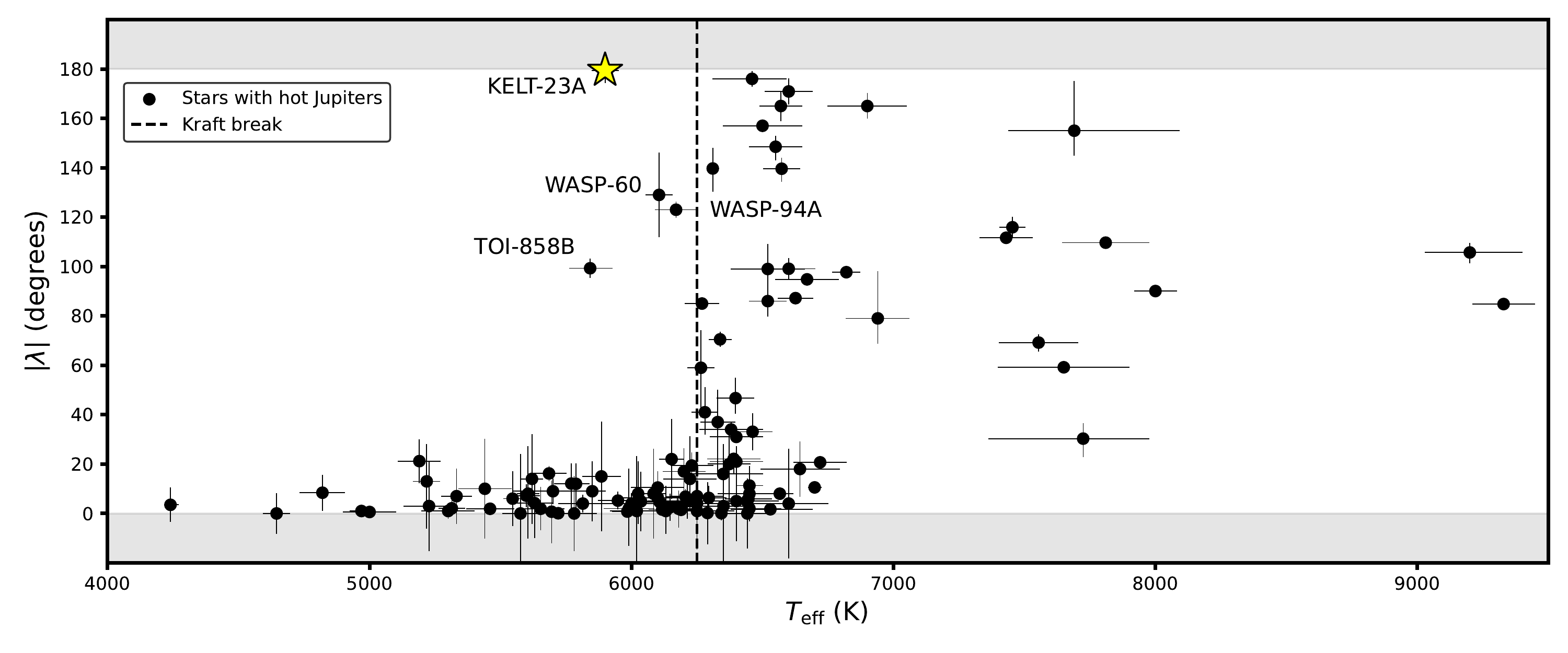}
    \caption{Sky-projected stellar obliquity ($\lambda$) versus stellar effective temperature ($T_{\rm eff}$) for stars with HJs (planets with $M_{\rm p} = 0.3 - 13 \, M_{\rm Jup}$ and $a/R_\star < 10$). The yellow star is KELT-23A and the dashed line is the Kraft break ($T_{\rm Kraft} \approx 6250$~K). Systems with $\lambda$ uncertainties greater than $30^\circ$ or with unreliable $\lambda$ measurements are excluded (see Section~\ref{sec:discussion}). KELT-23A is one of few cool stars with a HJ on a retrograde orbit ($|\lambda| > 90^\circ$). The three other cool stars with HJs on misaligned orbits are labeled: WASP-60 \citep{mancini2018wasp60}, WASP-94A \citep{neveu2014wasp94A, ahrer2024wasp94}, and TOI-858B \citep{hagelberg2023toi858B}. Like KELT-23A, the latter two of these systems have binary star companions, potentially suggesting a link between stellar multiplicity and spin-orbit misalignment. Data acquired from TEPCat on 1 May 2025 \citep{southworth2011tepcat}.}
    \label{fig:lambda_teff}
\end{figure*}

\begin{figure*}[t!]
    \centering
    \hspace{-0cm}\includegraphics[width=0.495\linewidth]{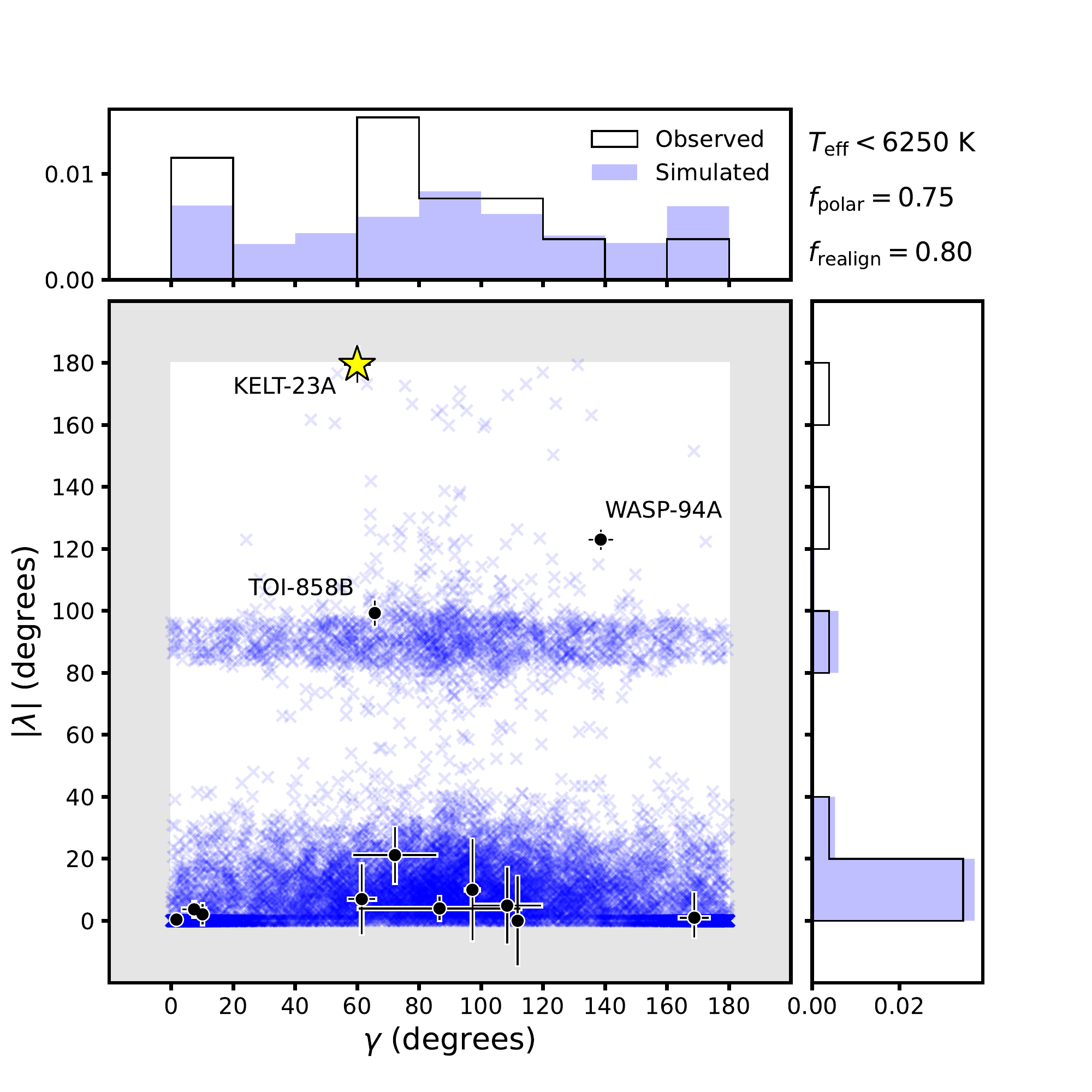}
    \includegraphics[width=0.495\linewidth]{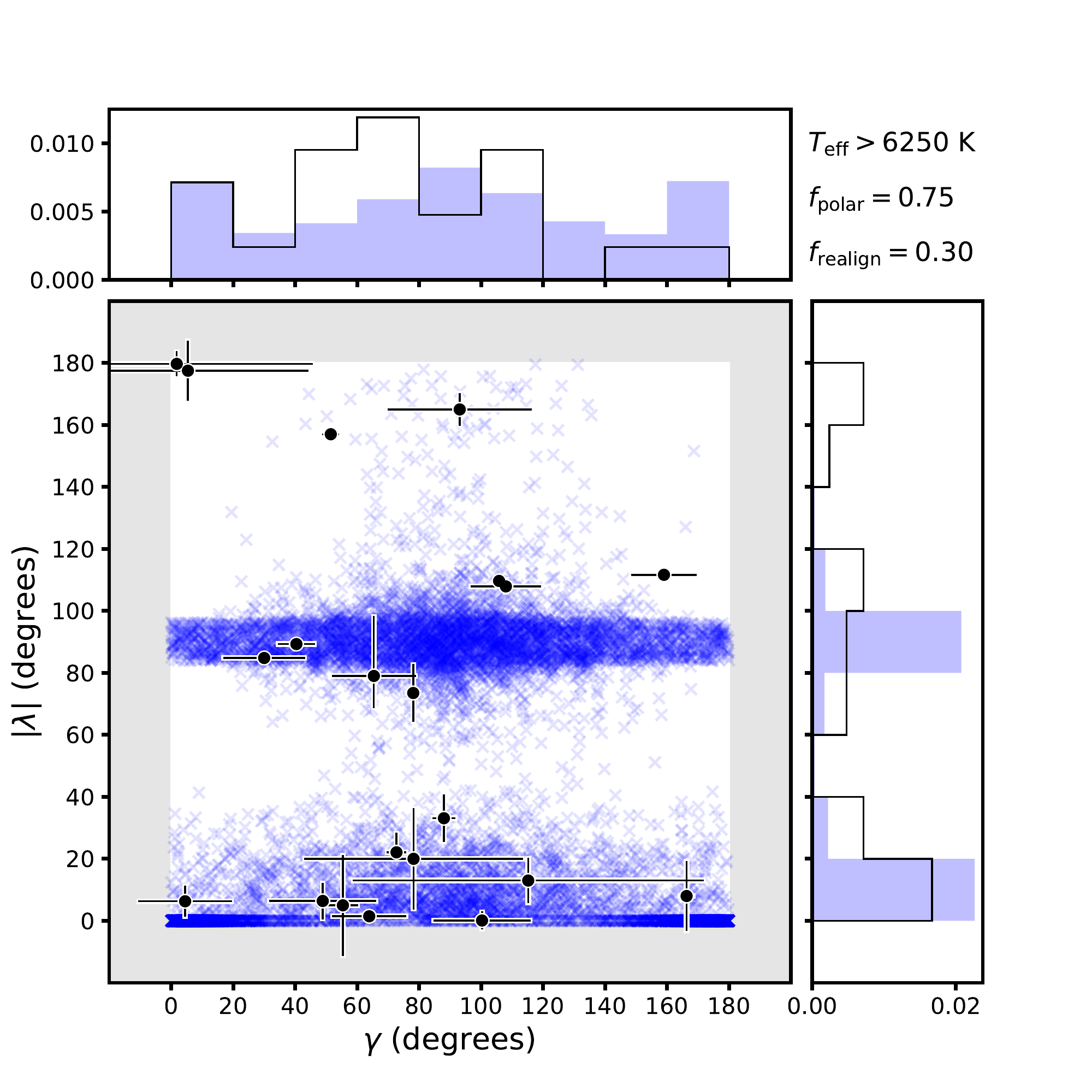}
    \caption{Sky-projected stellar obliquity ($\lambda$) versus sky-projected binary orbit angle ($\gamma$) for cool stars (left; $T_{\rm eff} < 6250$~K) and hot stars (right; $T_{\rm eff} > 6250$~K) with stellar binary companions and HJ companions ($M_{\rm p} = 0.3 - 13 \, M_{\rm Jup}$ and $a/R_\star < 10$). Black circles are the observed data and blue X's are simulated data from the toy model in Section~\ref{sec:popsynth}. Normalized histograms of the data are shown in the margins. Observed systems with $\lambda$ uncertainties greater than $30^\circ$ or with unreliable $\lambda$ measurements are excluded (see Section~\ref{sec:discussion}). Values of $\gamma$ for KELT-23A ($\gamma = 60 \pm 4^\circ$) and TOI-858B ($\gamma = 65.7 \pm 0.5^\circ$) were calculated using the approach in \citet{behmard2022companions}. For the remaining systems, we adopted the values of $\gamma$ from \citet{rice2024geometries}. Our toy model can reproduce the most salient features of the distribution, namely the excess of systems near $\gamma = 90^\circ$, $\gamma = 0^\circ$, and $\gamma = 180^\circ$, and the dearth of systems with $|\lambda|$ between $40^\circ$ and $80^\circ$.}
    \label{fig:lambda_gamma}
\end{figure*}

\section{Discussion} \label{sec:discussion}

We place KELT-23A into context with other stars with HJ companions (defined here as planets with $M_{\rm p} = 0.3 - 13 \, M_{\rm Jup}$ and $a/R_\star < 10$) in Figure~\ref{fig:lambda_teff}. This figure, which plots sky-projected stellar obliquity against stellar effective temperature, showcases the sharp transition between the mostly low obliquities of cool stars with HJs and the high obliquities of hot stars with HJs. First noted by \citet{winn2010obliquities} to occur near the Kraft break \citep{kraft1967break}, this transition is likely a consequence of different tidal realignment efficiencies in stars with different internal energy transport structures \citep{albrecht2022obluqities}. KELT-23A stands out as one of the only cool stars known to have a high stellar obliquity relative to its HJ companion, the other three being WASP-60 \citep[$|\lambda| = 129 \pm 17^\circ$;][]{mancini2018wasp60}, WASP-94A \citep[$|\lambda| = 123 \pm 3^\circ$;][]{neveu2014wasp94A, ahrer2024wasp94}, and TOI-858B \citep[$|\lambda| = 99.3^{+3.8}_{-3.7}$$^\circ$;][]{hagelberg2023toi858B}. We note that this figure excludes stars with $\lambda$ uncertainties greater than $30^\circ$. It also excludes the stars CoRoT-1, CoRoT-19, HATS-14, WASP-1, WASP-2, and WASP-23 due to their $\lambda$ measurements being unreliable (see \citealt{albrecht2022obluqities} for more details).

Interestingly, three of the four cool stars with HJs on misaligned orbits -- KELT-23A, WASP-94A, and TOI-858B -- have wide-separation stellar companions. KELT-23A has an M dwarf companion with a projected separation of 570~au, WASP-94A has a late F-type companion with a projected separation of 2700~au, and TOI-858B has a late F-type companion with a projected separation of 3000~au. This trend hints that outer companions play a role in misaligning  HJs around cool stars. One possible scenario involves the HJ beginning its life on an initially distant orbit that had its eccentricity and inclination excited via von Zeipel-Lidov-Kozai (ZLK) interactions with the wide-separation stellar companions \citep{vonzeipel1910, kozai1962secular, lidov1962evolution}, after which tidal interactions with the host star induced rapid orbital shrinking and circularization \citep{fabrycky2007kz}. Observational and theoretical studies have suggested that this mechanism may be responsible for an appreciable fraction (although likely no more than $\sim 50 \%$) of HJs \citep{naoz2012kz, dawson2015kz, petrovich2015kz, anderson2016kz, munoz2016kz, ngo2016friendsIV}. Star-planet ZLK interactions are also among the most favored mechanisms for producing planets on polar and retrograde orbits, making it a particularly compelling explanation for this subsample of the HJ population. In principle, it is also possible that the presence of binary companions triggered primordial spin-orbit misalignments due to nascent star-disk interactions \citep{batygin2012primordial, batygin2013misalignment, lai2014star-disc-binary, spalding2014misalignment, matsakos2017disks, zanazzi2018discwarping, zanazzi2018misalignments}.

Alternatively, these planets could have been driven inwards due to interactions with unknown planetary-mass companions with wide orbital separations \citep[e.g.,][]{chatterjee2008scattering, wu2011secular, beauge2012scattering, petrovich2015chem, petrovich2016planetkz, tayssendier2019secular}. Radial velocity surveys of single-star systems have found that nearly all stars with close-in Jupiter-mass companions have another planetary companion of equal or greater mass on a more distant orbit \citep{zink2023companions}. Some of these planet-planet interaction mechanisms are believed to be capable of producing HJs on retrograde orbits \citep{beauge2012scattering, lithwick2014secular, petrovich2016planetkz}. It is plausible that the presence of a wide stellar binary companion facilitates interactions between the two planets that ultimately drives one inwards \citep[e.g.,][]{yang2025hatp7}.

Regardless of how KELT-23A~b and similar planets arrived at their close-in orbits, we must also address the issue of how the planet has retained a retrograde orbit over a long timescale. While the exact tidal mechanisms responsible for spin-orbit realignment are still up for debate \citep{albrecht2022obluqities}, most predict that cool stars should achieve realignment on relatively short timescales \citep[e.g.,][]{winn2010obliquities, dawson2014tidal, zanazzi2024damping, zanazzi2025damping}. One possible explanation is that the planetary orbit has ``stalled'' at an antialigned or polar orientation. Using equilibrium tide and inertial wave dissipation models, \citet{xue2014evolution} and \citet{li2016tides} showed that this stalling can last many billions of years and usually coincides with a slowing of the stellar rotation rate, possibly explaining the low $v \sin{i_\star}$ of KELT-23A (also see \citealt{lai2012dissipation}). Lastly, we cannot rule out the possibility that KELT-23A~b arrived at its close-in orbit relatively recently and is currently in the process of having its orbit realigned with the stellar spin. Depending on the rate of spin-orbit realignment, it may be inevitable that some HJs orbiting cool stars will be caught in the act.

\begin{figure*}[t!]
    \centering
    \includegraphics[width=0.92\linewidth]{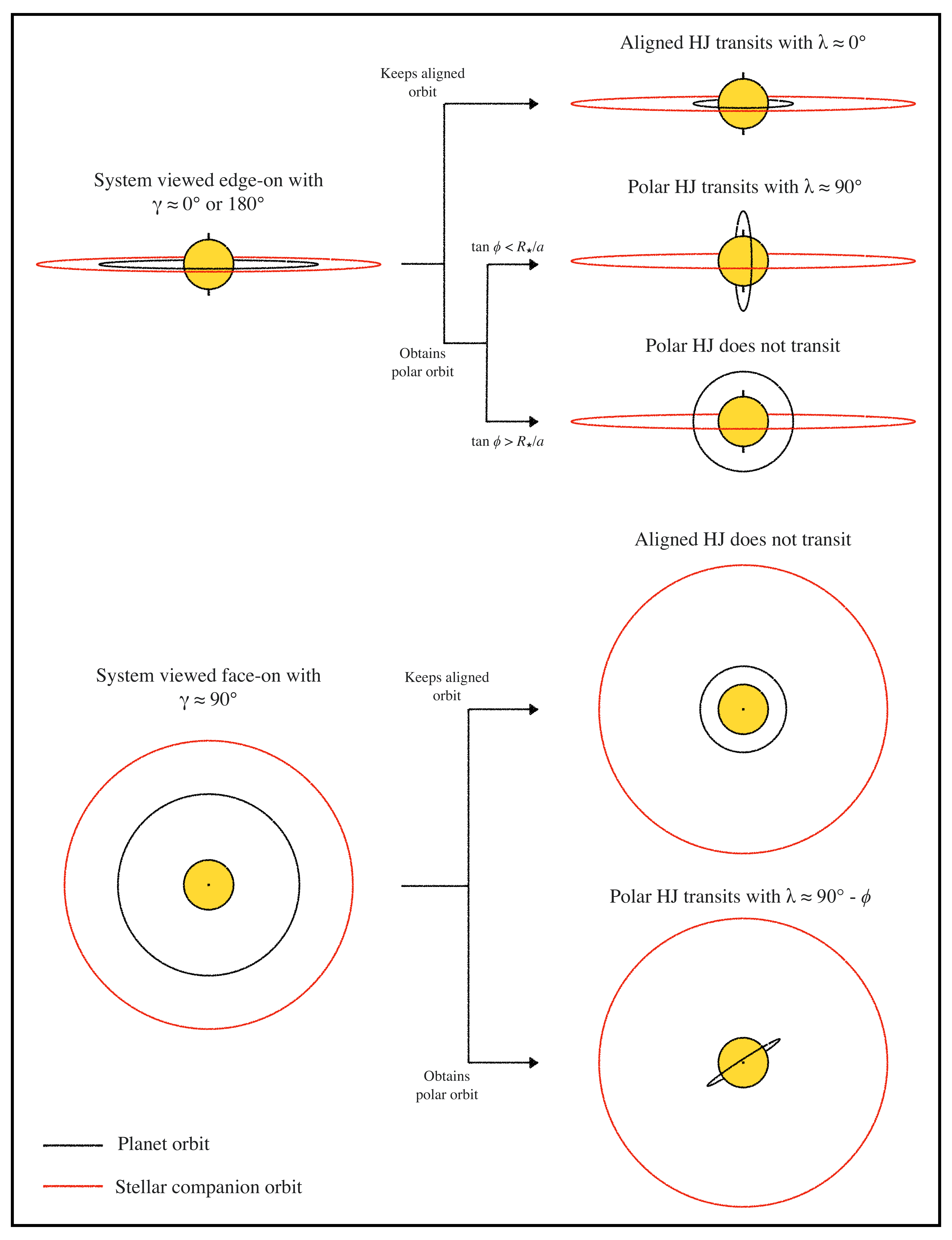}
    \caption{Visualizations of the two extreme cases of the toy model in Section~\ref{sec:popsynth}, in which the initial system is viewed edge-on (top) and face-on (bottom), and their possible outcomes. Initially edge-on systems generally produce transiting HJs with $\lambda \approx 0^\circ$ or $\lambda \approx 90^\circ$. Initially face-on systems only produce HJs that transit when the HJ is in a polar orbit. A wide range of $\lambda$ values are possible for these systems, where $\lambda \approx 90^\circ - \phi$ (note that $i_\star$ must be non-zero, else the R-M signal would be undetectable).}
    \label{fig:toymodel}
\end{figure*}

\subsection{Toy Model and Population Synthesis}\label{sec:popsynth}

To better understand the role stellar companions have on the production of HJs, we plotted $\lambda$ and $\gamma$ for cool stars ($T_{\rm eff} < 6250$~K) and hot stars ($T_{\rm eff} > 6250$~K) with stellar binary companions and HJ companions in Figure~\ref{fig:lambda_gamma}. Numerical simulations have suggested that these wide binary companions may play prominent roles in disrupting planetary systems, potentially producing HJs in the process \citep[e.g.,][]{kaib2013disruption}. Based on the (albeit small-numbered) distribution in Figure~\ref{fig:lambda_gamma}, there appears to be a preference for HJs in binary star systems to be found when the binary orbit is near face-on ($\gamma = 90^\circ$) or near edge-on ($\gamma = 0^\circ$ or $\gamma = 180^\circ$). We note that this pattern was first identified by \citet{behmard2022companions} and was later disputed by \citet{rice2024geometries} using a larger sample of systems that included non-HJ planets.


To determine whether this distribution can be reproduced via interactions between the planet (or the disk in which it formed) and the binary companion, we constructed a toy model and performed a population synthesis simulation of the resulting values of $\lambda$ and $\gamma$. The model, which is visualized in Figure~\ref{fig:toymodel}, is set up as follows:
\begin{enumerate}[itemsep=-0.1cm]
    \item Establish a system with a Jovian planet on a $1-10$~au orbit and a 0.5~$M_\odot$ stellar companion on a 1000~au orbit around a Solar-type star. Assume that the orbit of the planet and the stellar companion are approximately coplanar and aligned with the equator of the primary star, a configuration that recent papers have suggested may be common for systems with small, close-in planets \citep{christian2024aligned, rice2024geometries}.  Given that Jovian planets likely form in similar disks and that the orbits of warm Jupiters tend to be aligned with the equators of their host stars \citep{rice2022warm}, it is reasonable to extrapolate this trend to the cold Jupiters considered here.
    \item Sample initial orbital parameters for the system. Assume the possible orbital eccentricities of the stellar companion are uniformly distributed between 0 and 1,\footnote{In principle, very high eccentricities ($e \gtrsim 0.9$) can be excluded because the binary companions would disrupt the protoplanetary disks. In any case, we found that the results of the simulation are not strongly sensitive to variations of $e_{\rm max}$ between 0.8 and 1.0.} the cosine of the possible orbital inclinations are uniformly distributed between 0 and 1, and the possible arguments of periastron, arguments of ascending node, and true anomaly are uniformly distributed between 0 and $2\pi$. Transform the resulting orbital positions and motions into celestial coordinates and calculate $\gamma$.
    \item Assume the Jovian planet is driven inward via some unspecified mechanism and becomes a HJ with a 3-day orbital period, which corresponds to $a/R_\star = 8.75$ for a Solar-type star. Assume that the planet obtains a polar orbit with a frequency of $f_{\rm polar}$ and an aligned orbit with a frequency of $1 - f_{\rm polar}$ \citep[e.g.,][]{albrecht2021perpendicular}. If the planet obtains a polar orbit, randomly draw an angle $\phi$ from a uniform distribution between 0 and $\pi$, where $\phi$ is the rotation angle of the orbit normal about the stellar rotation axis (i.e., if the stellar spin axis points along the $y$ axis, $\phi$ describes where the orbit normal, which is always orthogonal to the $y$ axis for a polar orbit, points in the $x-z$ plane).
    \item If the HJ obtains a polar orbit, assume there is a probability of $f_{\rm realign}$ that the star realigns with the planet orbit before we observe it. If the star does realign, draw the final value of $\lambda$ from a half-Normal distribution with a mean of $0^\circ$ and a standard deviation of $15^\circ$.
    \item Determine if the resulting HJ transits and, if it does, calculate $\lambda$. In terms of the initial orbital inclination of the planet ($i_0$) and $\phi$, the condition for a polar HJ to transit is $\sin{i_0} \tan{\phi} < R_\star/a$. For simplicity, we assume $\lambda \approx 90^\circ - \phi$, which is accurate to within a few degrees.
    \item Repeat steps 1-5 $N$ times to generate a population.
\end{enumerate}
The result of this simulation for $N=10^5$ is shown in Figure~\ref{fig:lambda_gamma}. Based on visual inspection, we found that $f_{\rm polar} = 0.75$ with $f_{\rm realign} = 0.80$ most closely reproduces the observed distribution for cool stars and $f_{\rm polar} = 0.75$ with $f_{\rm realign} = 0.30$ most closely reproduces the observed distribution for hot stars. The toy model broadly reproduces the most salient features of the observed distributions, namely the preference for transiting HJs to be in systems with $\gamma$ near $0^\circ$, $90^\circ$, and $180^\circ$, in addition to the preference for values of $\lambda$ near $0^\circ$ and $90^\circ$ \citep{albrecht2021perpendicular}. The model is able to produce systems with $|\lambda|$ between $90^\circ$ and $180^\circ$, accounting for systems like KELT-23A and WASP-94A, but underpredicts the number of these systems we should observe. In addition, there may be an observational bias against systems with high $\lambda$, because these systems are more likely to have stars spinning face-on with lower values of $v \sin{i_\star}$ (and therefore lower R-M amplitudes), meaning the fraction of systems in these highly misaligned orientations may be even higher than the current data suggests. This implies that a non-negligible fraction of HJs arrive near their host stars with true retrograde, non-polar orbits and remain in those orientations for long enough to be observed.

It is interesting to compare our chosen value of $f_{\rm polar}$ to those predicted by different migration mechanisms. As mentioned previously, most mechanisms struggle to produce systems with polar or retrograde orbits, with planet-planet ZLK and planet-star ZLK being the most efficient \citep{anderson2016kz, petrovich2016planetkz}. However, these mechanisms are only predicted to produce planets on polar or retrograde orbits $25-50\%$ of the time, making them inconsistent with the results of our toy model. Primordial misalignment mechanisms involving interactions between the protoplanetary disk and the stellar companion may offer a more compelling explanation \citep{batygin2012primordial, batygin2013misalignment, lai2014star-disc-binary, spalding2014misalignment, matsakos2017disks, zanazzi2018discwarping, zanazzi2018misalignments}.

\section{Conclusion} \label{sec:conclusion}

Using the KPF spectrograph, we measured the sky-projected stellar obliquity of the Sun-like star KELT-23A, which hosts a transiting HJ companion with an orbital period of 2.26 days \citep{johns2019kelt23}. By jointly fitting the \textit{TESS} and KPF data of the star, we found $\lambda = 180.4^{+4.9}_{-4.7}$$^\circ$ and ($v \sin{i_\star} = 0.468^{+0.044}_{-0.043}$~km~s$^{-1}$), indicating a retrograde orbit and a slow sky-projected stellar rotational velocity. This low value of $v \sin{i_\star}$ may suggest that the stellar spin axis is being viewed close to face-on, which would mean the true obliquity is close to polar. This is one of a few stars with an effective temperature below the Kraft break that has a HJ on a misaligned orbit, providing an interesting case study of HJs orbiting cool stars that have not yet had their orbits realigned with the stellar spin by tides. In particular, the observed obliquity of KELT-23A may be consistent with realignment theories involving equilibrium tides and inertial wave dissipation, which predict that the orbits of HJs can stall at antialigned and polar orientations for Gyr timescales \citep{xue2014evolution, li2016tides}.

We explored the role that the wide-separation stellar companion KELT-23B may have played in creating a HJ with a retrograde orbit. When examining the sky-projected binary orbit angles ($\gamma$) of binary systems in which one star hosts a HJ with a $\lambda$ measurement, systems tend to clump near face-on binary orbits ($\gamma = 90^\circ$) and edge-on binary orbits ($\gamma = 0^\circ$ and $\gamma = 180^\circ$) \citep{behmard2022companions}. We found that this distribution can be broadly reproduced using a toy model in which a HJ begins its life on a distant orbit that is coplanar with both the equator of the primary star and the orbit of the wide-separation stellar companion, after which it migrates inwards and preferentially obtains an orbit that is either aligned with the stellar equator or polar.

\begin{acknowledgments}

We thank Heather A. Knutson and Joshua N. Winn for their helpful discussions involving this paper.

The data presented herein were obtained at the W. M. Keck Observatory, which is operated as a scientific partnership among the California Institute of Technology, the University of California, and the National Aeronautics and Space Administration. The Observatory was made possible by the generous financial support of the W. M. Keck Foundation. Keck Observatory occupies the summit of Maunakea, a place of significant ecological, cultural, and spiritual importance within the indigenous Hawaiian community. We understand and embrace our accountability to Maunakea and the indigenous Hawaiian community, and commit to our role in long-term mutual stewardship. We are most fortunate to have the opportunity to conduct observations from Maunakea.

We gratefully acknowledge the efforts and dedication of the Keck Observatory staff for support of KPF and remote observing.

S.G. is supported by an NSF Astronomy and Astrophysics Postdoctoral Fellowship under award AST-2303922. A.W.H. acknowledges funding support from NASA grant No. 80NSSC24K0161.

The research was carried out, in part, at the Jet Propulsion Laboratory, California Institute of Technology, under a contract with the National Aeronautics and Space Administration (80NM0018D0004)

\end{acknowledgments}





%
\facilities{TESS, Keck:I (KPF)}

\software{\texttt{BATMAN} \citep{kriedberg2015batman}, \texttt{DYNESTY} \citep{speagle2020dynesty}}

\bibliography{bibliography}{}
\bibliographystyle{aasjournalv7}

\end{document}